\documentclass[usenames,dvipsnames,12pt]{article}

\usepackage{graphicx}
\usepackage{amsmath}
\usepackage{amssymb}
\usepackage{latexsym}
\usepackage{mathrsfs}
\usepackage{amsthm}
\usepackage{setspace}
\usepackage{epsfig}
\usepackage{subfigure}
\usepackage{authblk}
\usepackage{amsfonts}
\usepackage{url}
\usepackage{float}
\usepackage{multibib}

\usepackage[authoryear,round]{natbib}	
\usepackage{mathtools}														 
\mathtoolsset{showonlyrefs=false}									 

\usepackage{geometry}
\usepackage[dvips]{color}
\newtheoremstyle{newplain}
{4pt}
{4pt}
{\itshape}
{}
{\itshape\bf}
{.}
{.5em}
{}
\theoremstyle{newplain}

\newtheoremstyle{newdefinition}
{4pt}
{4pt}
{}
{}
{\itshape\bf}
{.}
{.5em}
{}
\theoremstyle{newdefinition}

\newtheorem{vg}{Example}[section]

\newtheorem{remark}{Remark}[section]
\numberwithin{equation}{section}
\newcommand{\be}{\begin{equation}}
\newcommand{\ee}{\end{equation}}
\newcommand{\nee}{\nonumber\end{equation}}
\newcommand{\eel}[1]{\label{#1}\end{equation}}
\newcommand{\brmk}[1]{\begin{remark}\label{#1}\begin{em} }
\newcommand{\ermk}{ $\quad\triangleleft$\end{em}\end{remark}}
\newcommand{\bvg}[1]{\begin{vg}\label{#1}\begin{em} }
\newcommand{\evg}{ $\quad\triangleleft$\end{em}\end{vg}}

\usepackage{chngcntr}
\counterwithout{equation}{section}

\begin{document}
\bibliographystyle{plainnat}
\setlength{\abovedisplayskip}{8pt}
\setlength{\belowdisplayskip}{8pt}
\setlength{\abovedisplayshortskip}{4pt}
\setlength{\belowdisplayshortskip}{8pt}

\begin{titlepage}

\title{{\bf{\Huge Long Forward Probabilities, Recovery and the Term Structure of Bond Risk Premiums}}\thanks{This paper is based on research supported by the grants from the National Science Foundation CMMI-1536503 and DMS-1514698.}}
\author{Likuan Qin\thanks{likuanqin2012@u.northwestern.edu}}
\author{Vadim Linetsky\thanks{linetsky@iems.northwestern.edu}}
\author{Yutian Nie\thanks{ynie@u.northwestern.edu}}
\affil{\emph{Department of Industrial Engineering and Management Sciences}\\
\emph{McCormick School of Engineering and Applied Science}\\
\emph{Northwestern University}}
\date{}

\end{titlepage}

\maketitle
\begin{abstract}
We show that the martingale component in the long-term factorization of the stochastic discount factor due to  \citet{alvarez_2005using} and \citet{hansen_2009} is highly volatile, produces a downward-sloping term structure of bond Sharpe ratios, and implies that the long bond is far from growth optimality.  In contrast, the long forward probabilities forecast an upward sloping term structure of bond Sharpe ratios that starts from zero for short-term bonds and implies that the long bond is growth optimal.
Thus, transition independence and degeneracy of the martingale component are implausible  assumptions in the bond market.
\end{abstract}

%
%
%
%

\section{Introduction}

This paper extracts transitory and permanent (martingale) components in the long-term factorization of the stochastic discount factor (SDF) of \citet{alvarez_2005using} and \citet{hansen_2009} (see also  \citet{linetsky_2014long}). We posit an arbitrage-free dynamic term structure model (DTSM), estimate it on the time series of US Treasury yield curves, and explicitly determine the long-term factorization of the SDF via the Perron-Frobenius extraction of the principal eigenfunction following the methodology of \citet{hansen_2009} (see also \citet{linetsky_2014_cont}). The martingale component of the long-term factorization defines the {\em long-term risk-neutral probability measure} (\citet{hansen_2009}, \citet{hansen_2013}, \citet{borovicka_2014mis}) that can also be identified with the {\em long forward measure}, the long-term limit of $T$-maturity forward measures well-known in the fixed income literature (see \citet{linetsky_2014long} for details). Consistent with the calibrated structural example in \citet{borovicka_2014mis}, as well as the empirical literature relying on bounds and finite-maturity proxies for the long bond (\citet{alvarez_2005using}, \citet{bakshi_2012}, \citet{bakshia2015inquiry}), we find that the martingale component is highly volatile.

With the estimated long-term factorization in hand, we are able to empirically test the structural assumption of transition independence of the SDF underpinning the recovery result of \citet{ross_2011}. \citet{ross_2011} shows that under the assumptions that all uncertainty in the economy follows a discrete-time irreducible Markov chain and that the SDF process is transition independent, there exists a unique recovery of subjective transition probabilities of investors from observed Arrow-Debreu prices (\citet{carr_2012} extend to 1D diffusions on a bounded interval, \citet{walden_2013} extends to more general 1D diffusions, and \citet{linetsky_2014_cont} extend to general Markov processes). Under the assumption of rational expectations, it leads to the recovery of the data generating transition probabilities. Transition independence is the key assumption that allows Ross to appeal to the Perron-Frobenius theory to achieve a unique recovery. \citet{hansen_2013}, \citet{borovicka_2014mis}, \citet{martin_2013} and \citet{linetsky_2014_cont} connect Ross' recovery to the factorization of \citet{hansen_2009} and show that transition independence in a Markovian model implies that the martingale component in the long-term factorization of SDF is degenerate and equal to unity. \citet{hansen_2013} and \citet{borovicka_2014mis} point out that such degeneracy is inconsistent with many structural dynamic asset pricing models, as well as with the empirical evidence in \citet{alvarez_2005using} and \citet{bakshi_2012} based on bounds on the permanent and transitory components of the SDF.

In the present paper we directly extract the long-term factorization of the SDF and evaluate the magnitude of the martingale component in the US Treasury bond market and, as a consequence, evaluate the plausibility of the transition independence assumption in the bond market. First, we briefly recall the long-term factorization of the SDF (\citet{alvarez_2005using}, \citet{hansen_2009}, \citet{hansen_2012}, \citet{hansen_2013}, \citet{borovicka_2014mis}, \citet{linetsky_2014long}, \citet{linetsky_2014_cont}):
\be
\frac{S_{t+\tau}}{S_t}=\frac{1}{R^\infty_{t,t+\tau}}\frac{M_{t+\tau}}{M_t},
\ee
where $S_t$ is the pricing kernel process, $R^\infty_{t,t+\tau}$ is the gross holding period return on the {\em long bond} (limit of gross holding period returns $R_{t,t+\tau}^T=P_{t+\tau,T}/P_{t,T}$ on pure discount bonds maturing at time $T$ as $T$ grows asymptotically large), and $M_t$ is a martingale.
This martingale defines the long-term risk neutral (long forward) probability measure we denote by ${\mathbb L}$ (in this paper we denote the physical or data generating measure by ${\mathbb P}$ and the risk-neutral measure by ${\mathbb Q}$).
Under ${\mathbb L}$,
the long bond serves as the growth optimal numeraire portfolio (see Section 4.3 in \citet{borovicka_2014mis} and Theorem 4.2 in  \citet{linetsky_2014long}).  By Jensen's inequality, the expected log return on any other asset is dominated by the long bond:
\be
{\mathbb E}_t^{\mathbb L}\left[\log R_{t,t+\tau}\right]\leq {\mathbb E}_t^{\mathbb L}\left[\log R^\infty_{t,t+\tau} \right],
\ee
where $R_{t,t+\tau}=V_{t+\tau}/V_t$ is the gross holding period return on an asset with the value process $V$, and the expectation is taken under the long-term risk-neutral measure ${\mathbb L}$.
To put it another way,  only the covariance with the long bond is priced under ${\mathbb L}$, with all other risks neutralized by distorting the probability measure:
\be
{\mathbb E}_t^{\mathbb L}\left[R_{t,t+\tau}\right]-R^f_{t,t+\tau}=-{\rm cov}_t^{\mathbb L}\left(R_{t,t+\tau},\frac{1}{R^\infty_{t,t+\tau}}\right)R^f_{t,t+\tau},
\eel{covariance_price}
where $R^f_{t,t+\tau}=1/P_{t,t+\tau}$ is the gross holding period return on risk-free discount bond. Dividing both sides by the conditional volatility of the asset return $\sigma_t^{\mathbb L}(R_{t,t+\tau})$, the conditional Sharpe ratio under ${\mathbb L}$ is
\be
{\mathcal SR}_t^{\mathbb L}(R_{t,t+\tau})=-{\rm corr}_t^{\mathbb L}\left(R_{t,t+\tau},\frac{1}{R^\infty_{t,t+\tau}}\right)R^f_{t,t+\tau}\sigma_t^{\mathbb L}\left(1/R_{t,t+\tau}^\infty\right).
\eel{SR}
The perfect negative correlation then gives the \citet{hansen_1990implications} bound under ${\mathbb L}$:
\be
{\mathcal SR}_t^{\mathbb L}(R_{t,t+\tau})\leq \sigma_t^{\mathbb L}\left(1/R_{t,t+\tau}^\infty\right) R^f_{t,t+\tau}.
\eel{HJ}
For a more detailed presentation of the long forward measure ${\mathbb L}$ see \citet{linetsky_2014long}.

Assuming that the Markovian SDF is transition independent implies that the martingale component is degenerate, that is,
$S_{t+\tau}/S_t=1/R^\infty_{t,t+\tau}$.
This identifies ${\mathbb P}$ with ${\mathbb L}$, identifies
the long bond with the growth optimal numeraire portfolio in the economy (see also Result 5 in \citet{martin_2013} and Section 4.3 in \citet{borovicka_2014mis}), and implies that the only priced risk in the economy is the covariance with the long bond. In particular, applying Eq.\eqref{SR} to returns $R^T_{t,t+\tau}$ on pure discount bonds, Eq.\eqref{SR} predicts that bond Sharpe ratios are {\em increasing} in maturity and approach their upper bound (Hansen-Jagannathan bound \eqref{HJ}) at asymptotically long maturities.\footnote{While the long bond maximizes the expected log return, it does not generally maximize the Sharper ratio since ${\rm corr}_t^{\mathbb L}\left(R^\infty_{t,t+\tau},1/R^\infty_{t,t+\tau}\right)$ is not generally equal to $-1$. However, for sufficiently small holding periods this correlation is close to $-1$. In the empirical results in this paper, for three-month holding periods the empirically estimated ${\mathbb L}$-Sharpe ratio of the long bond is close to the upper bound given by the right hand side of equation Eq.\eqref{SR}, as discussed in Section 4.}
However, this sharply contradicts well known empirical evidence in the US Treasury bond market.

It is documented by  \citet{duffee2011sharpe}, \citet{frazzini2014betting} and \citet{van2015term} that short-maturity bonds have higher Sharpe ratios than long maturity bonds. \citet{backus2015term} and \citet{van2015term} provide recent bibliographies to the growing literature on the term structure of risk premiums. In this paper we focus on the term structure of bond risk premiums.
The empirical term structure of bond Sharpe ratios is generally downward sloping, rather than upward sloping.
\citet{frazzini2014betting} offer an explanation based on the leverage constraints faced by many bond market participants that result in their preference for longer maturity bonds over leveraged positions in shorter maturity bonds, even if the latter may offer higher Sharpe ratios.
Furthermore, empirical results in this paper show that leveraged short-maturity bonds achieve substantially higher expected log-returns than long-maturity bonds and, in particular, the (model-implied) long bond.
This empirical evidence puts in question the assumption of transition independent and degeneracy of the martingale component in the US Treasure bond market.

The rest of this paper is organized as follows. In Section 2 we estimate an arbitrage-free DTSM on the US Treasury bond data. There is an added challenge of the zero interest rate policy (ZIRP) in the US since December of 2008. Most conventional DTSM do not handle the zero lower bound (ZLB) well. Gaussian models allow unbounded negative rates, while CIR-type affine factor models feature vanishing volatility at the ZLB. Shadow rate models are essentially the only class of dynamic term structure models in the literature at present that are capable of handling the ZLB. The shadow rate idea is due to \citet{black_1995interest}. \citet{gorovoi_2004} provide an analytical solution for single-factor shadow rate models and calibrate them to the term structure of Japanese government bonds (JGB). \citet{kim2012term} estimate two-factor shadow rate models on the JGB data.
In this paper we estimate the two-factor shadow rate model B-QG2 (Black Quadratic Gaussian Two Factor) shown by Kim and Singleton to provide the best fit among the model specifications they consider in their investigation of the JGB market.

In Section \ref{s_lt_factor} we perform Perron-Frobenius extraction in the estimated model, extract the principal eigenvalue and eigenfunction, construct the long-term factorization of the pricing kernel, and recover the long-term risk neutral measure (long forward measure) dynamics of the underlying factors. We then directly compare market price of risk processes under the estimated data-generating probability measure and the recovered long-term risk-neutral measure.
The difference in these market prices of risk is identified with the instantaneous volatility of the martingale component. This difference is so large and, hence, the martingale component is so volatile that we reject the null hypothesis that the martingale is equal to unity (and, hence, the data-generating probability measure is identical to the long-term risk neutral measure) at the 99.99\% level. We note that our econometric approach in this paper is entirely different from the approaches of \citet{alvarez_2005using}, \citet{bakshi_2012} and \citet{bakshia2015inquiry} who rely on bounds on the transitory and martingale components, while we directly estimate a fully specified DTSM, explicitly accomplish the Perron-Frobenius extraction of \citet{hansen_2009} and obtain the permanent and martingale components in the framework of our DTSM.
We also note recent work by \citet{christensen2014nonparametric} who develops a non-parametric approach to the Perron-Frobenius extraction and estimates permanent and transitory components under structural (Epstein-Zin and power utility) specifications of the SDF calibrated to real per-capita consumption and real corporate earnings growth.
These three lines of inquiry, our parametric modeling and estimation based on asset market data, Christensen's modeling based on macro-economic fundamentals, and  \citet{alvarez_2005using}, \citet{bakshi_2012} and \citet{bakshia2015inquiry} approaches based on bounds,  are complementary and all result in the conclusion that the martingale component is highly economically significant.


In Section \ref{s_term_rp} we explore economic implications of our results. In Section 4.1, we use our model-implied long term bond dynamics to estimate expected log returns on the long bond and test how far it is from growth optimality implied by the assumption that the martingale component is unity. We find that duration-matched leveraged positions in short and intermediate maturity bonds have significantly higher expected log returns than long maturity bonds and, in particular, the long bond. We also estimate the realized term structure of Sharpe ratios for bonds of different maturities and conclude that it is downward-sloping,  consistent with the empirical evidence in \citet{duffee2011sharpe} and \citet{frazzini2014betting}.  We further consider Sharpe ratio forecasts under our estimated probability measures ${\mathbb P}$ and ${\mathbb L}$.
We find, in particular, that ${\mathbb L}$ implies forecasts for excess returns on shorter-maturity bonds (up to three years) that are essentially zero (risk-neutral), while  significant excess returns with high Sharpe ratios are observed empirically in this segment of the bond market and correctly forecast by our estimated ${\mathbb P}$ measure. Thus, identifying ${\mathbb P}$ and ${\mathbb L}$ leads to sharply distorted risk-return trade-offs in the bond market.
Finally, in Section \ref{s_lift_forecast} we show that using the ${\mathbb L}$ measure to forecast the expected timing of the Federal Reserve policy lift-off implied by the term structure of interest rates yields a forecast that is virtually indistinguishable from the risk-neutral forecast, while forecasting under the ${\mathbb P}$ measure yields a substantially different forecast.

\section{Dynamic Term Structure Model Estimation}
\label{s_model_estimate}
\begin{figure}[H]
\centering
\includegraphics[width=70mm,height=65mm]{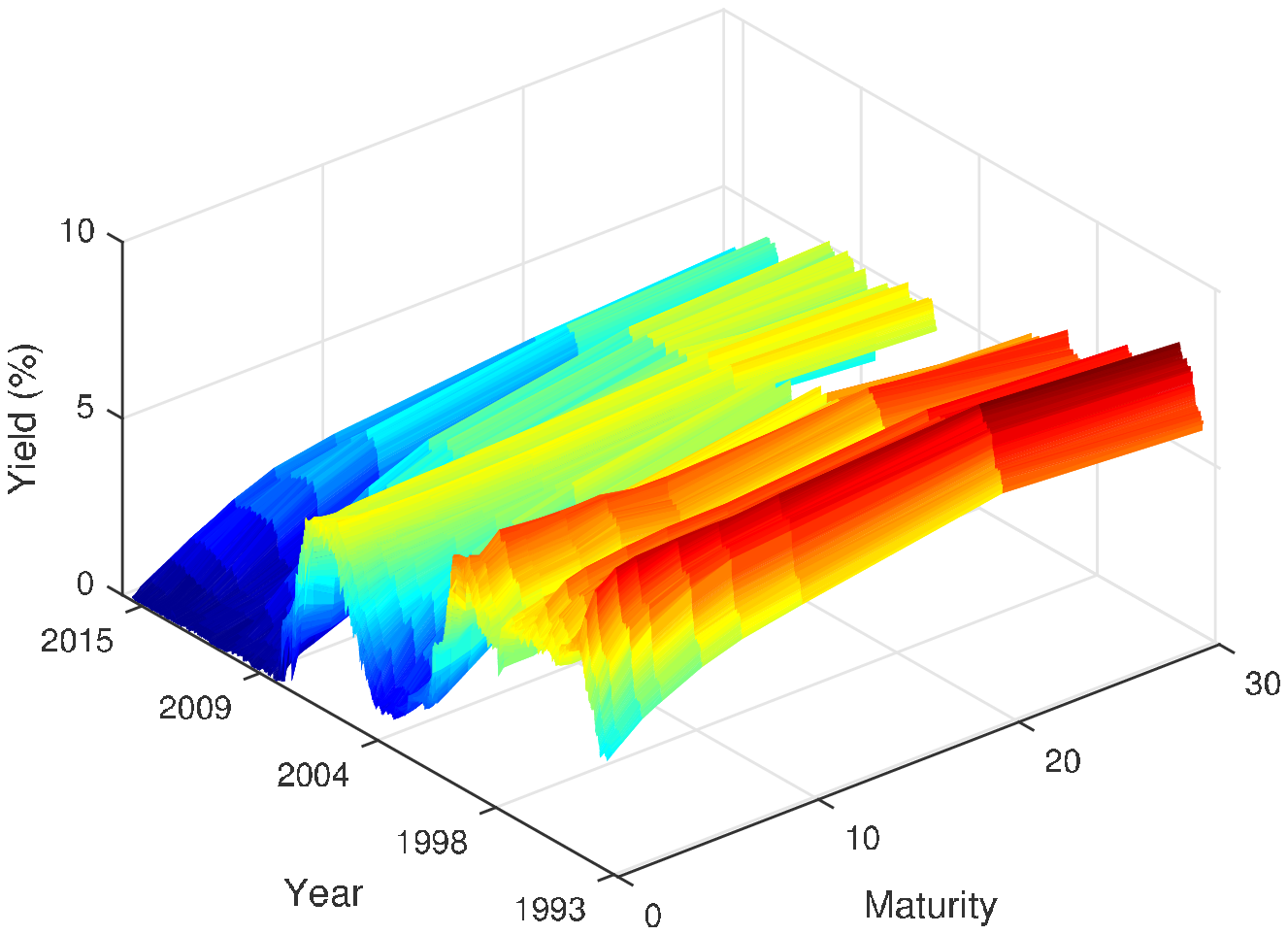}
\includegraphics[width=80mm,height=65mm]{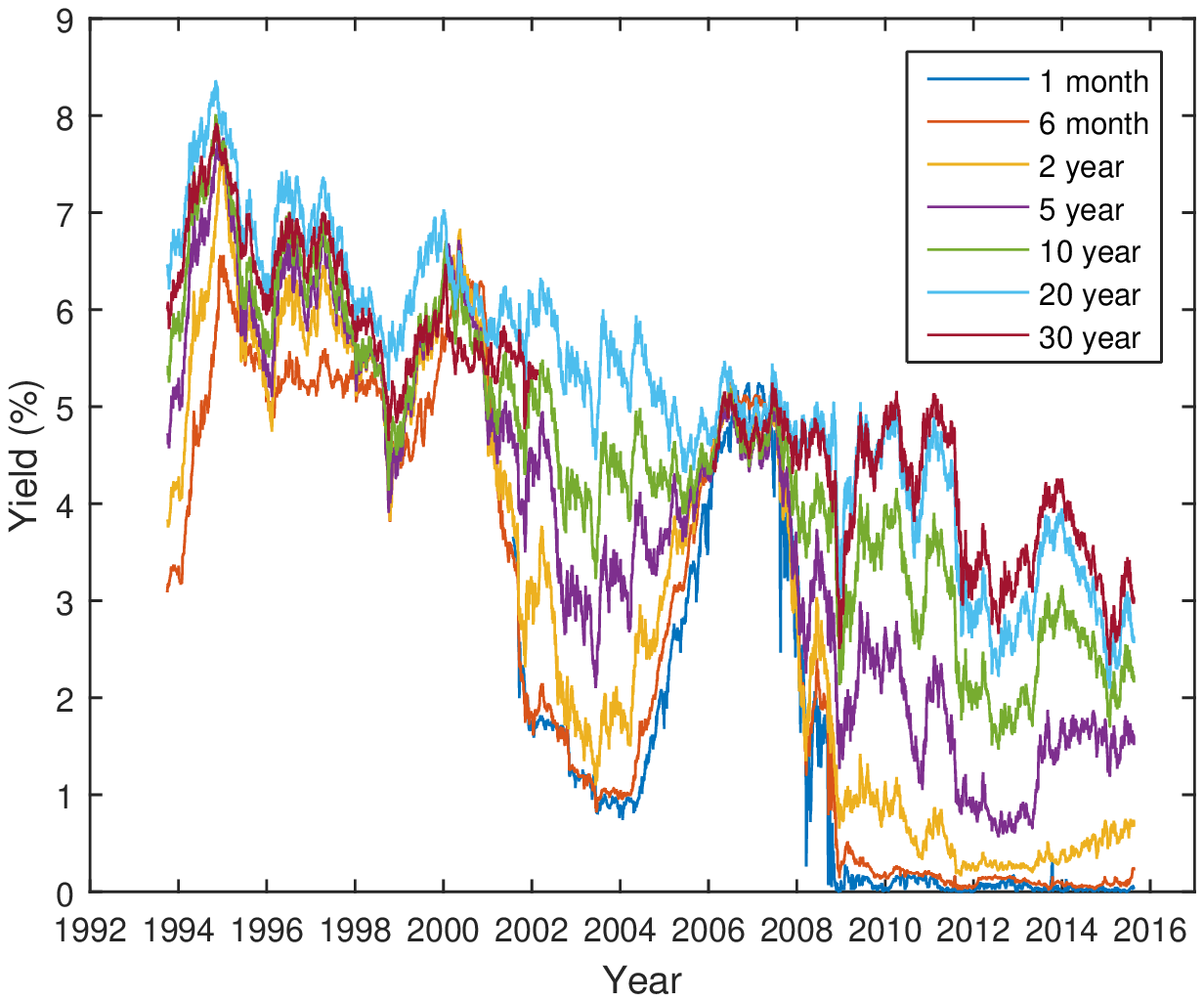}
\caption{US Treasury zero-coupon yield curves bootstrapped from CMT yield curves.}
\label{data}
\end{figure}

We use the data set of daily constant maturity (CMT) US Treasury bond yields from 1993-10-01 to 2015-08-19 available from the Federal Reserve Economic Data (FRED) web site (the same data are available from the US Treasury web site and are published daily by the Federal Reserve Board in the H.15 daily releases). The data include daily yields for Treasury constant maturities of 1, 3 and 6 months, and 1, 2, 3, 5, 7, 10, 20 and 30 years. Since our focus is on the Perron-Frobenius extraction of the principal eigenvalue and eigenfunction governing the long-term factorization, we include the long end of the yield curve with 20 and 30 year maturities. We choose 1993-10-01 as the start date of our data set because the 20 year maturity is available starting from this date. We observe that, while the yield curve is typically upward sloping between 10 and 20 years, on many dates it is nearly flat or slightly downward sloping between 20 and 30 year maturities.
Thirty year yield data are missing over the 4-year period from 2002-02-19 to 2006-02-08. One month yield data are missing over the 8-year period from 1993-10-01 to 2001-07-30, where the data start with three month yields. These missing data do not pose any challenges to our estimation procedure.
We obtain zero-coupon yield curves from CMT yield curves via cubic splines bootstrap. Figure \ref{data} shows our time series of bootstrapped zero-coupon yield curves.

We assume that the state of the economy is governed by a two-factor continuous-time Gaussian diffusion under the data-generating probability measure
$\mathbb{P}$:
\be
dX_t=K^\mathbb{P}(\theta^\mathbb{P}-X_t) dt+\Sigma dB_t^\mathbb{P},
\ee
where $X_t$ is a two-dimensional (column) vector, $B_t^\mathbb{P}$ is a two-dimensional standard Brownian motion, $\theta^\mathbb{P}$ is a two-dimensional vector, and $K^\mathbb{P}$ and $\Sigma$ are $2\times2$ matrices.
We assume an affine market price of risk specification
$\lambda^\mathbb{P}(X_t)=\lambda_0^\mathbb{P}+\Lambda^\mathbb{P} X_t,$
where $\lambda_0^\mathbb{P}$ is a two-dimensional vector and $\Lambda^\mathbb{P}$ is a 2x2 matrix,
so that $X_t$ remains Gaussian under the risk-neutral probability measure $\mathbb{Q}$:
\be
dX_t=K^\mathbb{Q}(\theta^\mathbb{Q}-X_t) dt+\Sigma dB_t^\mathbb{Q},
\ee
where $K^\mathbb{Q}=K^\mathbb{P}+\Sigma\Lambda^\mathbb{P}$ and $K^\mathbb{Q}\theta^\mathbb{Q}=K^\mathbb{P}\theta^\mathbb{P}-\Sigma\lambda_0^\mathbb{P}$.

To handle the ZIRP since December of 2008, we follow  \citet{kim2012term} and specify \citet{black_1995interest} shadow rate as the shifted quadratic form of the Gaussian state vector, and the nominal short rate as its positive part (here $'$ denotes matrix transposition and $(x)^+=\max(x,0)$:
\be
r(X_t)=(\rho+\delta' X_t+ X_t' \Phi X_t)^+.
\ee
This is the B-QG2 (Black-Quadratic Gaussian two-factor) specification of \citet{kim2012term}.
Following \citet{kim2012term}, we impose the following conditions to achieve identification: $K^\mathbb{P}_{12}=0, \delta=0, \Sigma=0.1 I_2$, where $I_2$ is the $2\times2$ identity matrix.
To ensure existence of the long-term limit (see \citet{linetsky_2014_cont}), we impose two additional restrictions. We require that the eigenvalues of $K^\mathbb{P}$ have positive real parts, and $\Phi$ is positive semi-definite. The first restriction ensures that $X$ is mean-reverting under the data-generating measure $\mathbb{P}$ and possesses a stationary distribution. The second restriction ensures that the short rate does not vanish in the long run. The mode of the short rate under the stationary distribution is $(\rho+(\theta^\mathbb{P})'\Phi\theta^\mathbb{P})^+$. If $\Phi$ is not positive semi-definite, the mode of the short rate under the stationary distribution can be zero.
We decompose
\be
\Phi=\begin{bmatrix} 1&0\\ A & 1\end{bmatrix} \begin{bmatrix} D_{1}&0\\0 & D_{2}\end{bmatrix}\begin{bmatrix} 1& A \\0 & 1\end{bmatrix},
\ee
and require that $D_{1},D_{2}\geq0$ and $D_1 D_2>0$.

Due to the positive part in the short rate specification, in contrast to one-factor shadow rate models that admit analytical solutions  (\citet{gorovoi_2004}), the two-factor model does not possess an analytic solution for bond prices.
Consider the time-$t$ price of the zero-coupon bond with maturity at time $t+\tau$ and unit face value:
\be
P(\tau,X_t)=\mathbb{E}^\mathbb{P}_t[e^{-\int_t^{t+\tau} r(X_s) ds}].
\ee
Since the state process is time-homogeneous Markov, the bond pricing function $P(\tau,x)$
satisfies the pricing PDE
\be
\frac{\partial P}{\partial \tau}-\frac{1}{2}\text{tr}(\Sigma\Sigma'\frac{\partial^2 P}{\partial x\partial x'} )-\frac{\partial P'}{\partial x} K^\mathbb{Q}(\theta^\mathbb{Q}-x)+r(x)P=0
\ee
with the initial condition $P(0,x)=1$.
We compute
bond prices by solving the PDE numerically via an operator splitting finite-difference scheme as in  Appendix A of \citet{kim2012term}.

Our estimation strategy follows \citet{kim2012term}. Observed bond yields $Y^O_{t,\tau_i}$ are assumed to equal their model-implied counterparts $Y_{t,\tau_i}=Y(\tau_i,X_t)=-(1/\tau_i)\log P(\tau_i,X_t)$ plus mutually and serially independent Gaussian measurement errors $e_{t,\tau_i}$.
The model is estimated using the extended Kalman-filter based quasi-maximum likelihood function. We follow \citet{kim2013estimation} in estimating standard errors using the approach of \citet{bollerslev1992quasi}.
Parameter estimates and standard errors are given in Table \ref{parameter_estimate}. Average pricing errors are given in Table \ref{fit_error}.
Our pricing errors are slightly higher than those reported by \citet{kim2012term}, where the model is estimated on weekly JGB data. It is not surprising, since we use daily data for all maturities from 1 month to 30 years, whereas \citet{kim2012term} use weekly data with JGB maturities up to 10 years.

\begin{table}[H]
\begin{tabular}{ccc}
\hline
$K^\mathbb{Q}$ & 0.3220 (0.0032)& 0.0415 (0.0005)\\
& 0.6391 (0.0073)& 0.0809 (0.0017)\\
$\theta^\mathbb{Q}$ & 0.9302 (0.0138) & \\
& -5.9261 (0.0727) &\\
$\gamma$ & -0.0048 (0.0002) & \\
$D_1$ & 0.2723 (0.0090) &\\
$D_2$ & 0.0223 (0.0007) &\\
$A$ & 0.3238 (0.0066) & \\
$\lambda^\mathbb{P}_a$ & -0.8929 (0.0556) &\\
& -0.9589 (0.0347) &\\
$\Lambda^\mathbb{P}_b$ & -3.3292 (0.8822) & 0.4152 (0.005)\\
& 4.2136 (1.1461) & 0.4012 (0.0997)\\
\hline
\end{tabular}
\caption{Model parameter estimates and standard errors (in parenthesis).}
\label{parameter_estimate}
\end{table}

\begin{table}[H]
\begin{tabular}{c c c c c c c c c c c}
\hline
 1m &3m&6m&1yr&2yr&3yr & 5yr &7yr &10yr & 20yr &30yr\\ \hline
13 & 12 & 8 & 8 & 14 & 13 & 10 & 9 & 12 & 15 & 13\\ \hline
\end{tabular}
\caption{Average pricing errors (in basis points).}
\label{fit_error}
\end{table}

\begin{figure}[H]
\centering
\includegraphics[width=75mm,height=65mm]{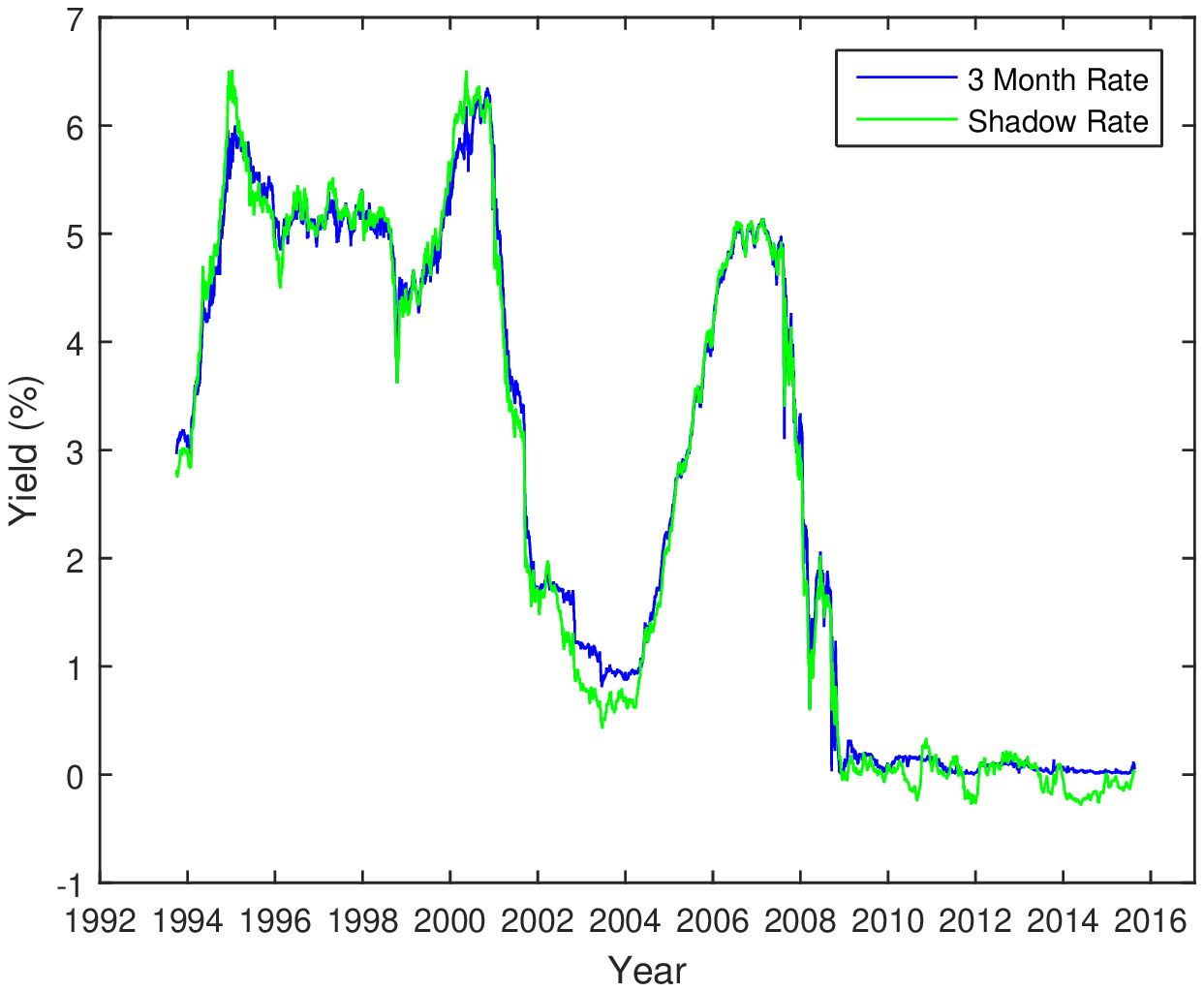}
\includegraphics[width=75mm,height=65mm]{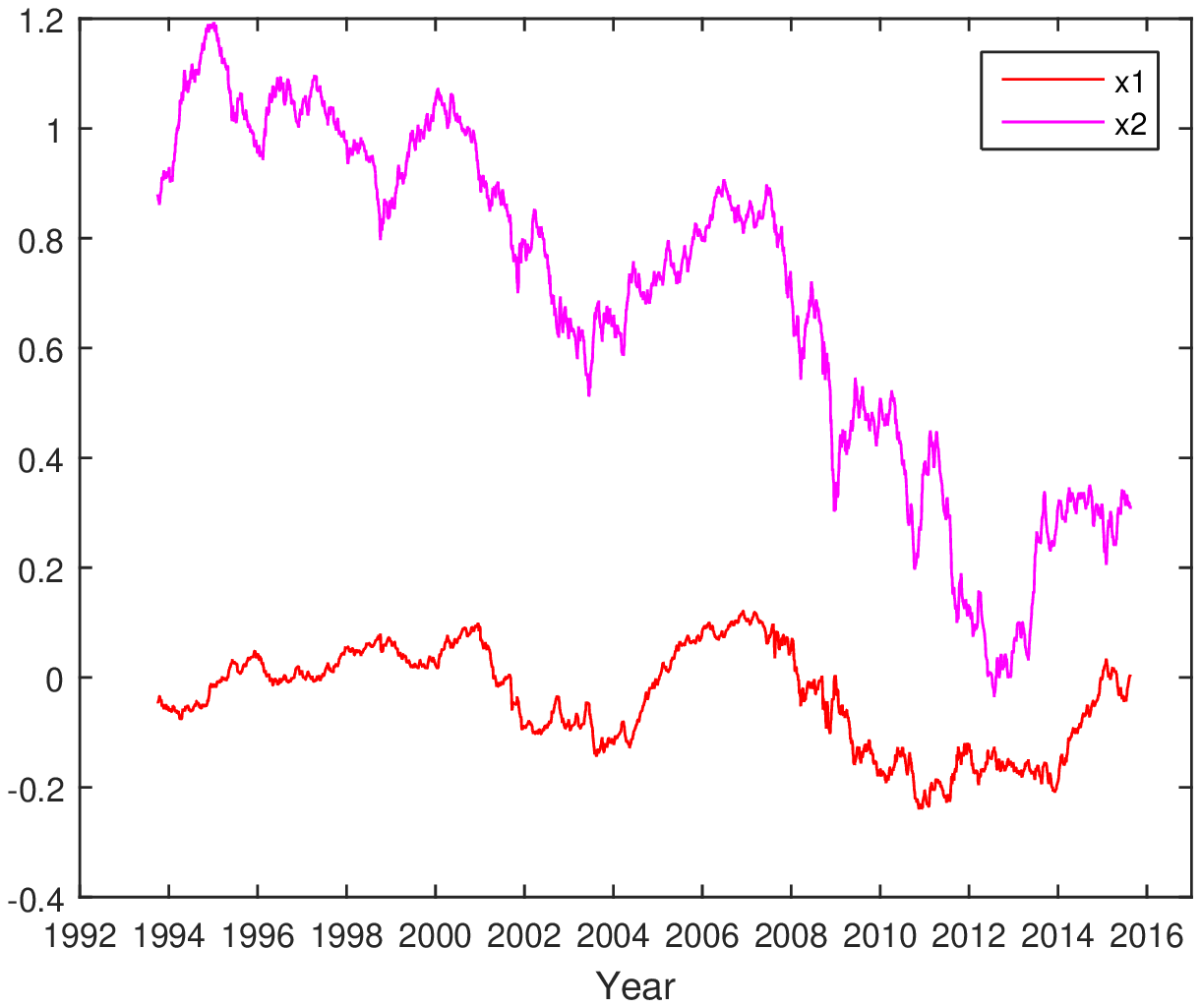}
\caption{Filtered paths of the state variables and the model-implied shadow rate.}
\label{g_state_path}
\end{figure}

\section{Long-Term Factorization}
\label{s_lt_factor}

We now turn to constructing the long-term factorization of the SDF process
\be
S_t=e^{-\int_0^t r(X_s)ds}e^{-\int_0^t \lambda^{\mathbb P}(X_s)dB_s^{\mathbb P}-\frac{1}{2}\int_0^t \|\lambda^{\mathbb P}(X_s)\|^2ds}
\ee
in the estimated dynamic term structure model.  Consider the gross holding period return on the zero-coupon bond with maturity at time $T$ over the period from $s$ to $s+t$, $R^T_{s,s+t}=P(T-s-t,X_{s+t})/P(T-s,X_s)$. We are interested in the limit as $T$ goes to infinity (holding period return on the zero-coupon bond of asymptotically long maturity). In Markovian models, if the long-term limit exists (see \citet{linetsky_2014long} for sufficient conditions and mathematical details), then
\be
\lim_{T\rightarrow \infty}R^T_{s,s+t}=e^{\lambda t}\frac{\pi(X_{s+t})}{\pi(X_s)}
\ee
for some $\lambda$ and a positive function $\pi(x)$, with $\pi(x)$ serving as the positive (principal) eigenfunction of the (time-homogeneous Markovian) pricing operator  with the eigenvalue $e^{-\lambda t}$:
\be
{\mathbb E}^{\mathbb P}_0[S_t \pi(X_t)]= e^{-\lambda t}\pi(X_0),
\ee
where $S_t$ is the SDF. For the sake of brevity, here we do not repeat the theory of long-term factorization and its connection to the Perron-Frobenius theory and refer the reader to \citet{hansen_2009}, \citet{hansen_2012}, \citet{borovicka_2014mis}, \citet{linetsky_2014_cont} and \citet{linetsky_2014long}.

\begin{figure}[H]
\centering
\includegraphics[width=90mm,height=65mm]{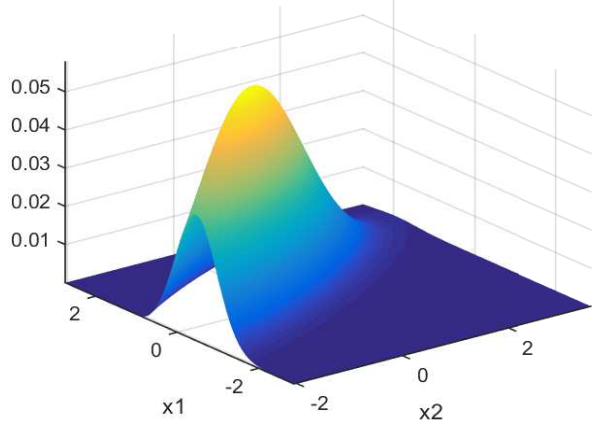}
\caption{Principal eigenfunction $\pi(x_1,x_2)$ with eigenvalue $\lambda=0.0282$.}
\label{eigenfunction}
\end{figure}

In the framework of our model the bond pricing function $P(t,x)$ is determined numerically by solving the bond pricing PDE by finite differences. We also determine the principal eigenfunction $\pi(x)$ numerically as follows. Choosing some error tolerance $\epsilon$, we solve the bond pricing PDE for an increasing sequence of times to maturity indexed by integers $n$, consider the ratios $P(n+1,x)/P(n,x)$ as $n$ increases, and stop at $n=N$ such that $M_N-m_N \leq \epsilon$ for the first time, where $M_n=\max_{x\in \Omega} P(n+1,x)/P(n,x)$ and $m_n=\min_{x\in \Omega} P(n+1,x)/P(n,x)$ and the max and min are computed over the grid in the domain $\Omega$ where we approximate the bond pricing function by the computed numerical solution of the PDE. The eigenvalue and the principal eigenfunction are then approximately given by $e^{-\lambda} = (m_N+M_N)/2$ and $\pi(x)=e^{\lambda N}P(N,x)$ in the domain $x\in\Omega$ (with the error tolerance $\epsilon$).
Figure \ref{eigenfunction} plots the computed eigenfunction $\pi(x)$. The corresponding principal eigenvalue is $\lambda=0.0282$. While there is no exact analytical solution for the eigenfunction in this shadow rate model due to the presence of the positive part function in the nominal short rate, this numerically determined eigenfunction is well approximated by an exponential-quadratic function of the form
\be
\pi(x)\approx e^{-1.92x_1^2-0.62x_2^2+1.69x_1 x_2 +1.62x_1-0.96x_2}
\eel{eq_exp_quad_approx}
on the domain $[-0.3, 0.2]\times[-0.1, 1.2]$ of values containing the filtered paths of the state variables, similar to quadratic term structure models (QTSM) (see \citet{linetsky_2014_cont} for details on positive eigenfunctions in ATSM and QTSM).

With the principal  eigenfunction $\pi(x)$ and eigenvalue $\lambda$ in hand, we explicitly obtain the long-term factorization:
\be
S_t = \frac{1}{L_t}M_t, \quad L_t = e^{\lambda t}\frac{\pi(X_t)}{\pi(X_0)}, \quad M_t = S_t e^{\lambda t}\frac{\pi(X_t)}{\pi(X_0)},
\eel{martingale_component}
where $L_t = R^\infty_{0,t}$ is the long bond process (gross return from time zero to time $t$ on the zero-coupon bond of asymptotically long maturity) determining the transitory component $1/L_t$, and $M_t$ is the martingale (permanent) component of the long-term factorization. In particular, we can now recover the ${\mathbb L}$ measure by applying Girsanov's theorem.
First, applying It\^{o}'s formula to $\log\pi(x)$ and using the SDE for $X$ under ${\mathbb P}$ we can write:
\be
\log \frac{\pi(X_t)}{\pi(X_0)}= \int_0^t \frac{\partial \log \pi}{\partial x'}(X_s) \Sigma dB_s^{\mathbb P}
+ \int_0^t \left(\frac{1}{2}\text{tr}(\Sigma\Sigma'\frac{\partial^2 \log\pi}{\partial x\partial x'})(X_s)+ \frac{\partial \log \pi}{\partial x'}(X_s) b^\mathbb{P}(X_s)  \right) ds,
\ee
where $b^{\mathbb P}(x)=K^{\mathbb P}(\theta^{\mathbb P}-x)$ is the drift under the data-generating measure.
Next, we recall that the eigenfunction satisfies the (elliptic) PDE (without the time derivative):
\be
\frac{1}{2}\text{tr}(\Sigma\Sigma'\frac{\partial^2 \pi}{\partial x\partial x'})(x)+\frac{\partial \pi}{\partial x'}(x) b^\mathbb{Q}(x)+(\lambda-r(x))\pi(x)=0,
\ee
where $b^{\mathbb Q}(x)=K^{\mathbb Q}(\theta^{\mathbb Q}-x)$ is the drift under the risk-neutral measure.
Using the identity $\frac{\partial^2 \log\pi}{\partial x\partial x'}=\frac{1}{\pi}\frac{\partial^2 \pi}{\partial x\partial x'}-\frac{1}{\pi^2}\frac{\partial \pi}{\partial x} \frac{\partial \pi}{\partial x'}$ and the PDE, we can write:
\be
\begin{split}
&\log \frac{\pi(X_t)}{\pi(X_0)}=\int_0^t (\frac{1}{\pi}\frac{\partial \pi}{\partial x'})(X_s) \Sigma dB_s^{\mathbb P}\\
&
+ \int_0^t \left(r(X_s)-\lambda-(\frac{1}{2\pi^2}\frac{\partial \pi}{\partial x'}\Sigma\Sigma' \frac{\partial \pi}{\partial x})(X_s)+(\frac{1}{\pi}\frac{\partial\pi}{\partial x'}\Sigma\lambda^\mathbb{P})(X_s)\right)ds.
\end{split}
\ee
Substituting this into the expression in Eq.\eqref{martingale_component} for the martingale $M_t$, we obtain:
\be
M_t=e^{-\int_0^t v_s dB^{\mathbb P}_s-\frac{1}{2}\int_0^t \|v_s\|^2ds}
\ee
with the instantaneous volatility process:
\be
v_t=\lambda^{\mathbb P}(X_t)-\lambda^{\mathbb L}(X_t),
\ee
where $\lambda^{\mathbb P}(x)$ is the drift of the state vector under the data-generating measure ${\mathbb P}$, and we introduced the following notation
\be
\lambda^{\mathbb L}(x):=\frac{1}{\pi(x)}\Sigma'\frac{\partial \pi}{\partial x}(x).
\ee
The martingale defines the long-term risk neutral measure ${\mathbb L}$. Applying Girsanov's theorem, we obtain the drift of the state vector $X$ under ${\mathbb L}$:
\be
b^{\mathbb L}(x)=b^{\mathbb Q}(x)+\Sigma\lambda^{\mathbb L}(x),
\eel{eq_L_drift}
where $\lambda^{\mathbb L}(X_t)$ is thus identified with the market price of risk process under the long-term risk-neutral measure ${\mathbb L}$.
The instantaneous volatility $v_t=v(X_t)$ of the martingale component is equal to the difference between the market prices of risk under the data-generating measure ${\mathbb P}$ and the long-term risk neutral measure ${\mathbb L}$ and is explicitly expressed in terms of the principal eigenfunction:
\be
v_t=\lambda^{\mathbb P}(X_t)-\frac{1}{\pi(X_t)}\Sigma'\frac{\partial \pi}{\partial x}(X_t).
\ee

Using the exponential-quadratic approximation for the principal eigenfunction \eqref{eq_exp_quad_approx}, we obtain an affine approximation for the market price of risk under the long-term risk neutral measure $\mathbb{L}$:
\be
\lambda^{\mathbb{L}}(x)\approx\begin{bmatrix}
0.162\\
-0.096\\
\end{bmatrix}+\begin{bmatrix}
-0.383 & 0.169 \\
0.169 & -0.124\\
\end{bmatrix}\begin{bmatrix}
x_1\\
x_2\\
\end{bmatrix}.
\eel{eq_L_MRP}
Substituting it into the expression for the drift of the state variables under ${\mathbb L}$ \eqref{eq_L_drift}, we obtain a Gaussian approximation for the dynamics of the state variables under $\mathbb{L}$.

We can now explicitly compare the data-generating and long-term risk-neutral dynamics. By inspection we see that all the parameters entering the  market prices of risk under $\mathbb{L}$ \eqref{eq_L_MRP} are significantly smaller in magnitude than the parameters in the market prices of risk under the data-generating measure $\mathbb{P}$:
\be
\lambda^{\mathbb{P}}(x)=\begin{bmatrix}
-0.8929\\
-0.9589\\
\end{bmatrix}+\begin{bmatrix}
-3.3292 & 0.4152 \\
4.2136 & 0.4012\\
\end{bmatrix}\begin{bmatrix}
x_1\\
x_2\\
\end{bmatrix}.
\ee
Thus, we obtain the instantaneous volatility of the martingale component as a function of the state:
\be
v(x)\approx\begin{bmatrix}
-1.055\\
-0.863\\
\end{bmatrix}+\begin{bmatrix}
-2.946 & 0.246 \\
4.045 & 0.525\\
\end{bmatrix}\begin{bmatrix}
x_1\\
x_2\\
\end{bmatrix}.
\ee

We now test the null hypothesis $\mathbb{P}=\mathbb{L}$ (equivalently, degeneracy of the martingale component, $v_t=0$).
The market price of risk under ${\mathbb P}$ contains five independent parameters ($\Lambda_{12}$ is fixed in terms of the risk-neutral parameters due to our identification condition $K^\mathbb{P}_{12}=0$) that are estimated with standard errors given in Table 1. The market price of risk parameters under the long-term risk-neutral measure are uniquely determined (recovered) from the risk-neutral parameters without any additional errors (over and above the errors in estimating the risk-neutral parameters, which are generally substantially smaller than the errors in estimating the market prices of risk under the data-generating measure).
Taking the risk-neutral parameters as given, we thus approximate asymptotic standard errors of our estimated parameters of the volatility of the martingale component $v_i(x)=v_i+\sum_j v_{ij}x_j$, $v_i=\lambda^\mathbb{P}_{i}-\lambda^\mathbb{L}_{i}$ and $v_{ij}=\Lambda^\mathbb{P}_{ij}-\Lambda^\mathbb{L}_{ij}$, with our estimated standard errors of market price of risk parameters (estimated in Table 1 following the approach of \citet{bollerslev1992quasi}). We then compute the $p$-values for each of the five null hypothesis $v_1=0$, $v_2=0$, $v_{11}=0$, $v_{21}=0$, $v_{22}=0$ (recall that $\lambda_{12}^{\mathbb P}$ is fixed by our identification condition). The $p$-values for the null hypothesis $v_1=0$, $v_2=0$ and $v_{22}=0$ are $0.0000$ computed to four decimals. The $p$-values for $v_{11}=0$ and $v_{21}=0$ are $0.0008$ and $0.0004$, respectively.
Thus, the null hypothesis that $v_t=0$ (the martingale component is unity, and the long-term risk-neutral measure is identified with the data-generating measure) is rejected at the 99.99\% level.

\section{The Term Structure of Bond Risk Premiums}





\label{s_term_rp}
We now turn to the empirical examination of the term structure of bond risk premiums.
Table \ref{sharpe} displays realized average quarterly excess returns, standard deviations and Sharpe ratios for zero-coupon bonds of maturities from one to thirty years, as well as the model-implied long bond, over the period from 1993-10-01 to 2002-02-15 and from 2006-02-09 to 2015-08-19 when the 30-year bond data are available. Excess holding period returns are computed over the three-month zero-coupon bond yields known at the beginning of each quarter. We observe that the term structure of Sharpe ratios is downward sloping, with the one-year bond earning the quarterly Sharpe ratio of 0.49 -- about two and a half times the Sharpe ratio of the zero-coupon 30-year bond over the same period. These Sharpe ratios are computed from the raw data and, as such, are model independent. The quarterly Sharpe ratio of the model-implied long bond is 0.15 -- slightly lower than the realized Sharpe ratio of the 30-year bond. This shape of the term structure of Sharpe ratios is in broad agreement with the findings of \citet{duffee2011sharpe} and \citet{frazzini2014betting} and is incompatible with the increasing term structure of Sharpe ratios arising under the assumption of transition independence and degeneracy of the martingale component in the long-term factorization.

\begin{table}[H]
\begin{tabular}{c|c c c c c c c c c}
\hline
Maturity &1 &2&3&5&7&10&20&30&LB\\
\hline
Exc. Ret. &$0.17\%$&$0.37\%$&$0.50\%$&$0.79\%$&$1.03\%$&$1.20\%$&$2.18\%$&$2.34\%$&$2.39\%$\\
St. Dev. &$0.35\%$&$0.90\%$&$1.46\%$&$2.54\%$&$3.46\%$&$4.75\%$&$8.19\%$&$12.08\%$&$16.33\%$\\
Sharpe &0.49 &0.41&0.34&0.31&0.30&0.25&0.27&0.19&0.15\\
\hline
\end{tabular}
\caption{Realized average quarterly excess returns, standard deviations and Sharpe ratios for zero-coupon bonds of maturities from one to thirty years and the model-implied long bond (LB) over the period from 1993-10-01 to 2002-02-15 and from 2006-02-09 to 2015-08-19 when the 30-year bond data are available. Excess returns are computed over the three-month zero-coupon bond yield known at the beginning of each quarter.}
\label{sharpe}
\end{table}

\begin{figure}[H]
\centering
\includegraphics[width=75mm,height=65mm]{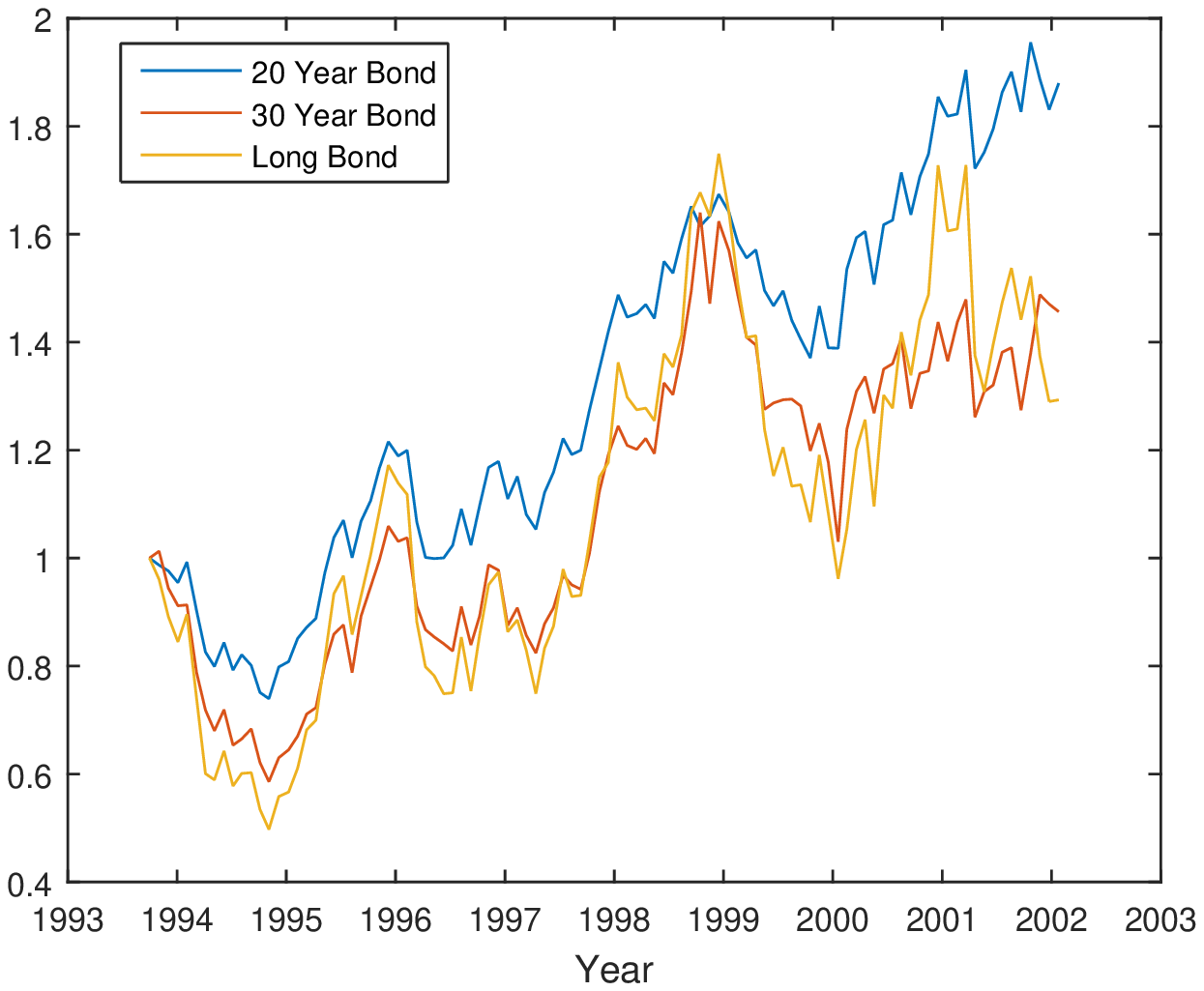}
\includegraphics[width=75mm,height=65mm]{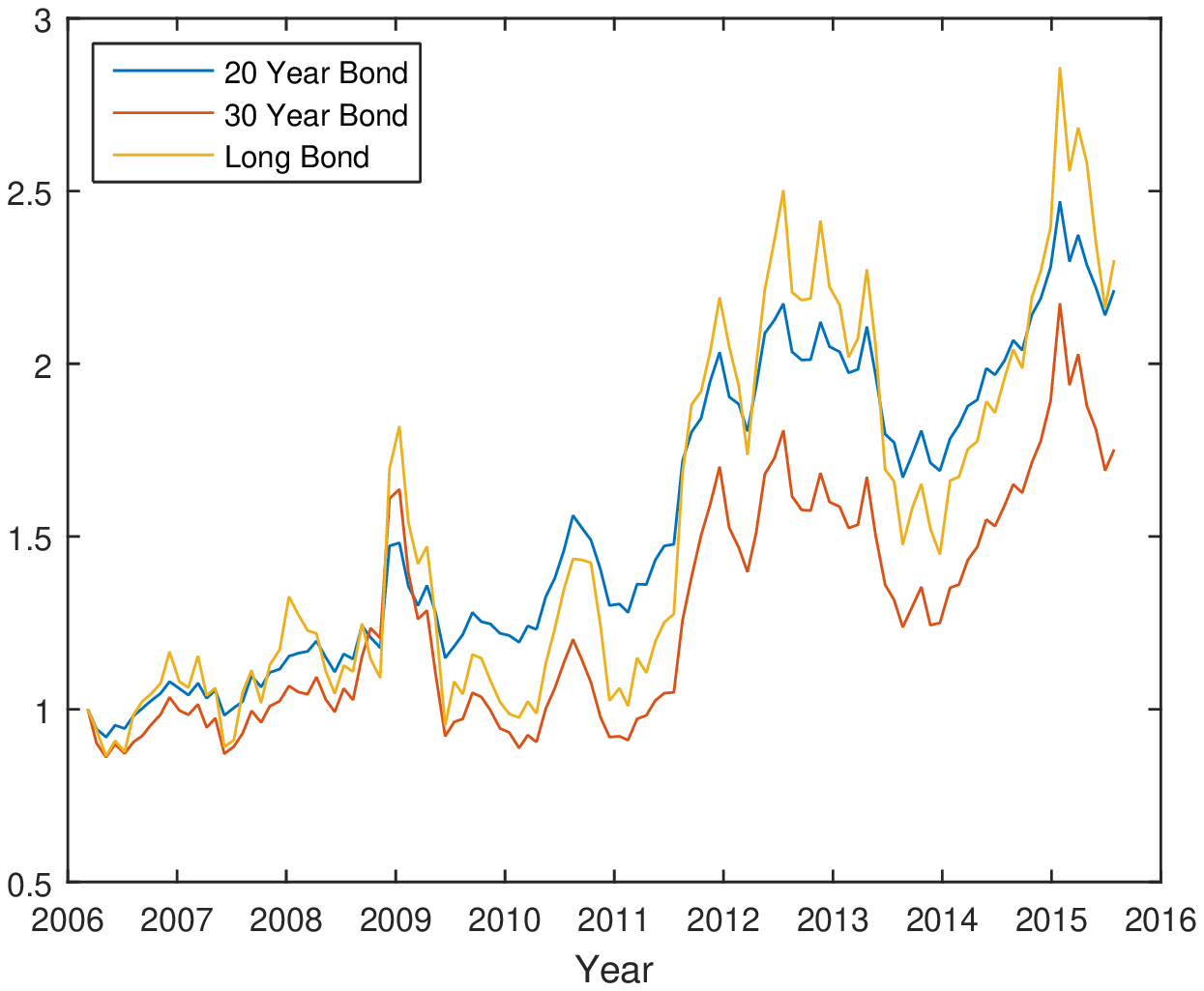}
\caption{Wealth processes investing in 20- and 30-year zero-coupon constant maturity bonds and the long bond.}
\label{long_bond_path}
\end{figure}

The model-implied long bond quantities are computed as follows. Recall that the long bond gross return process is given by $L_t=R_{0,t}^\infty=e^{\lambda t}\pi(X_t)/\pi(X_0)$.
Figure \ref{long_bond_path} displays the model-implied path of the long bond in our estimated DTSM obtained by evaluating the expression $e^{\lambda t}\pi(X_t)/\pi(X_0)$ on the filtered path of the state vector $X_t$ given in Figure \ref{g_state_path}, where the principal eigenfunction and eigenvalue are given in Figure 3. The figure also displays the wealth (gross return) processes of investing in 20- and 30-year  constant maturity zero-coupon bonds for comparison.
The time series is separated into two sub-periods since the 30 year bond was discontinued in 2002 and resumed in 2006.
Specifically, the 20-year time series shows the value over time of the initial investment of one dollar in the 20-year zero-coupon bond rolled over at three month intervals back into the 20-year bond.

In the previous literature researchers used 20- to 30-year bonds as proxies for the long bond.
In our framework of the fully specified DTSM, we have access to the model-implied long bond dynamics and can use it as a model-based proxy for the unobservable long bond. Figure 4 shows that during the first period from 1993 to 2003 the model-implied long bond was closer to the 30-year bond, while during the second period from 2006 to 2015 it was closer to the 20-year bond. However, during each of the two sub-periods the model-implied long bond path is appreciably distinct from the 20- and 30-year bonds.

Table \ref{log10_2} displays average realized quarterly log-returns for duration-matched leveraged or de-leveraged investments in zero-coupon bonds of different maturities that match
the duration of the ten- and twenty-year bond over the period from 1993-10-01 to 2002-02-15 and from 2006-02-09 to 2015-08-19 when the 30-year bond data are available.
We observe that leveraged investments in shorter-maturity bonds produce significantly higher average log-returns than duration-matched de-leveraged investments in longer maturity bonds. Using our model-implied long bond time series displayed in Figure \ref{long_bond_path}, we estimate the average expected log-return on the long bond to equal 1.98\% over this period. Comparing this with the data in Table \ref{log10_2}, we see that all of the leveraged investments in bonds of maturities from one- to ten-years leveraged to the twenty-year duration produce significantly higher average log-returns. The un-leveraged investment in twenty-year bonds also produces a substantially higher average log-return.
Leveraged investments in one- to seven-year bonds leveraged to match ten year duration also produce average log-returns higher than the long bond.
These results strongly reject growth optimality of the long bond, consistent with the high volatility of the martingale component in the long-term factorization established in Section \ref{s_lt_factor}.

\begin{table}[H]
\begin{tabular}{c|c c c c c c c c}
\hline
Maturity (years) &1 &2&3&5&7&10&20&30\\
\hline
Log-ret.  (10y dur.) &$2.34\%$ &$2.46\%$&$2.27\%$&$2.15\%$&$2.05\%$&$1.80\%$&$1.72\%$&$1.42\%$\\
\hline
Log-ret. (20y dur.) &$3.83\%$ &$3.99\%$&$3.58\%$&$3.34\%$&$3.15\%$&$2.67\%$&$2.55\%$&$1.96\%$\\
\hline
\end{tabular}
\caption{Realized average quarterly log-returns for leveraged (and de-leveraged) investing in zero-coupon bonds of different maturities matched to ten and twenty year durations. For the ten-year duration-matched strategies, for maturities from one to seven years the investment is leveraged by borrowing at the three-month rate  to match 10-year duration, and de-leveraged for 20 and 30 year maturities to match the 10 year duration. The period from 1993-10-01 to 2002-02-15 and from 2006-02-09 to 2015-08-19 when the 30-year bond data are available.}
\label{log10_2}
\end{table}


We next compare model-based conditional forecasts of excess returns, volatility and Sharpe ratios of zero-coupon bonds of different maturities under the data-generating measure ${\mathbb P}$ estimated in Section \ref{s_model_estimate} and the long-term risk-neutral measure ${\mathbb L}$ obtained via Perron-Forbenius extracton in Section \ref{s_lt_factor}.
Table \ref{bond_SR_pred_avg} displays average conditional excess return, volatility and Sharpe ratio forecasts under ${\mathbb P}$ and ${\mathbb L}$.
Reported values are obtained by calculating conditional forecasts along the filtered sample path of the state vector $X_t$ given in Figure 2 and taking the averages over the time period. Excess return forecasts are over the 3-month zero-coupon bond yield known at the beginning of each quarter.
Sharpe ratio forecasts are computed as the ratios of excess return forecast to the volatility forecast. Comparing Sharpe ratio forecasts in Table \ref{bond_SR_pred_avg} with Table \ref{sharpe}, we observe that $\mathbb{P}$-measure Sharpe ratio forecasts exhibit the downward-sloping term structure broadly comparable with the downward-sloping term structure of realized Sharpe ratios in Table \ref{sharpe}. In contrast, the ${\mathbb L}$-measure forecasts exhibit a generally upward-sloping term structure that starts near zero for one- to three-year maturities (${\mathbb L}$-measure forecasts are essentially risk-neutral for these shorter maturities) and increases towards  the Hansen-Jagannathan bound in Eq.\eqref{HJ} discussed in the Introduction.
The bound is approximately attained by the long bond. While the long bond is growth optimal, it does not generally maximize the Sharper ratio since ${\rm corr}_t^{\mathbb L}\left(R^\infty_{t,t+1},1/R^\infty_{t,t+1}\right)$ is not generally equal to $-1$. However, for sufficiently small holding periods this correlation is close to $-1$. Indeed, in Table \ref{bond_SR_pred_avg} compare the empirically estimated average quarterly ${\mathbb L}$-Sharpe ratio of the long bond of 0.18 with its average quarterly volatility also equal 0.18.

%
%

\begin{table}[H]
\begin{tabular}{|c c | c c c c c c c|}
\hline
& & 1yr & 3yr & 5yr & 10yr & 20yr &30yr & Long Bond\\ \hline
$\mathbb{P}$ &\text{Ex. Ret.} &$0.16\%$& $0.45\%$ & $0.60\%$ & $0.70\%$ & $0.67\%$& $0.63\%$ & $0.58\%$\\
& \text{St. Dev.} & $0.40\%$ & $1.03\%$& $1.59\%$& $3.99\%$& $9.36\%$& $12.95\%$ & $16.98\%$\\
& \text{Sharpe} & $0.40$ & $0.44$& $0.38$& $0.18$& $0.07$& $0.05$ & $0.03$\\ \hline
$\mathbb{L}$ &\text{Ex. Ret.} &$-0.02\%$& $0.02\%$ & $0.17\%$ & $0.71\%$ & $1.75\%$& $2.43\%$ & $3.19\%$\\
& \text{St. Dev.} & $0.42\%$ & $1.06\%$& $1.61\%$& $4.09\%$& $9.77\%$& $13.63\%$ & $18.04\%$\\
& \text{Sharpe} & $-0.05$ & $0.02$& $0.11$& $0.17$& $0.18$& $0.18$ & $0.18$\\\hline
\end{tabular}
\caption{Average conditional 3-month excess return, volatility and Sharpe ratio  ${\mathbb P}$- and ${\mathbb L}$-forecasts for zero-coupon bonds of maturities from one to thirty years and the model-implied long bond (LB) over the period from 1993-10-01 to 2002-02-15 and from 2006-02-09 to 2015-08-19 when the 30-year bond data are available. Excess return forecasts are over the 3-month zero-coupon bond yield known at the beginning of each quarter.}
\label{bond_SR_pred_avg}
\end{table}


\section{Forecasting the ZIRP Lift-off}
\label{s_lift_forecast}

\begin{figure}[H]
\centering
\includegraphics[width=75mm,height=65mm]{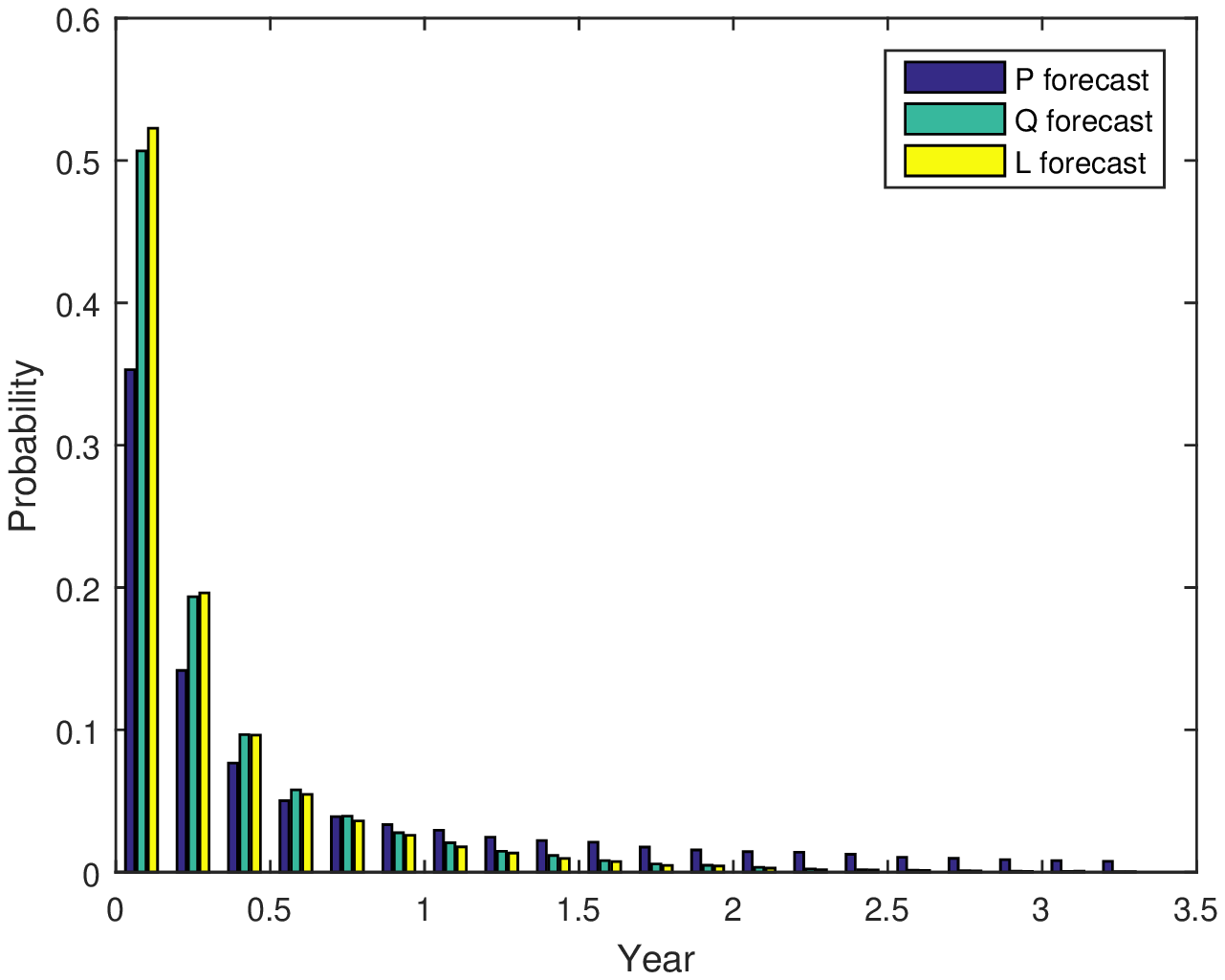}
\includegraphics[width=75mm,height=65mm]{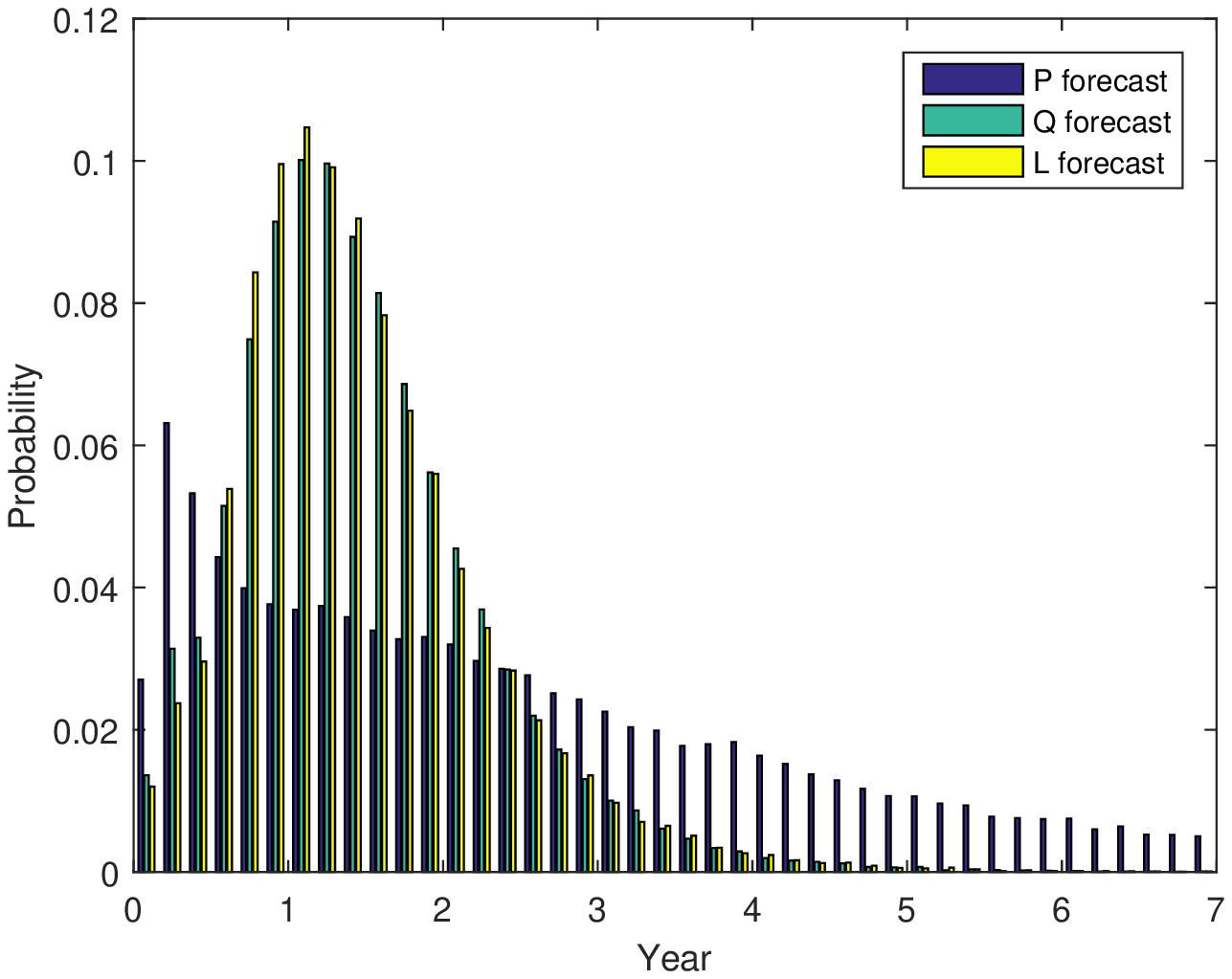}
\caption{Distribution of ZIRP lift-off time under $\mathbb{P}$, $\mathbb{Q}$ and $\mathbb{L}$ as of Aug. 19, 2015 (left) and Dec. 30, 2011 (right)}
\label{rate_lift}
\end{figure}

We next compare ${\mathbb P}$- and ${\mathbb L}$-forecasts of the timing of the Federal Reserve's zero interest rate policy lift-off. Specifically, we apply our estimated DTSM to simulate the first passage time of the short rate above 25 bps  from below as of August 19, 2015 (the last day in our data set) under ${\mathbb P}$, ${\mathbb L}$ and ${\mathbb Q}$.
Figure \ref{rate_lift} displays the simulated distributions of the first passage time.
Table \ref{rate_lift1} displays the mean and median of ${\mathbb P}$-, $\mathbb{Q}$- and ${\mathbb L}$-distributions. We observe that $\mathbb{Q}$ and $\mathbb{L}$ produce forecasts that are virtually indistinguishable, while $\mathbb{P}$ produces a significantly different forecast, and the first passage time distribution has a substantially heavier right tail. This is consistent with our previous result in Section 4 that the long-term risk-neutral measure $\mathbb{L}$ is very close to the risk-neutral measure $\mathbb{Q}$ when forecasting expectations computed over time horizons up to several years.
In this case the support of the distribution of the first passage time is concentrated primarily over the period up to three years. Using ${\mathbb L}$ over such time horizons in the bond market is essentially indistinguishable from using ${\mathbb Q}$. We also show the forecast as of December 30, 2011 to illustrate an earlier date during the ZIRP period with flatter term structure and more negative estimated shadow rate (note that this forecast is subject to look ahead bias since our DTSM parameters are estimated based on the time series over the entire period).
In this example the expected time of sitting at the zero bound is much longer. Again, while the ${\mathbb P}$-mean forecast is just under three years (cf. the actual lift-off in December of 2015 -- four years), the ${\mathbb L}$-mean forecast is half as long at about a year and a half and is very close to the risk-neutral forecast. In both cases, the ${\mathbb P}$-forecasts have a fat right tail corresponding to the possibility of ``secular stagnation" scenarios of sitting at the zero bound for a long time, while the ${\mathbb L}$- and ${\mathbb Q}$-forecasts have substantially thinner right tails and do not put much probability on those scenarios. While the ${\mathbb P}$-forecasts appear economically plausible in these examples, the point of these examples is not to discuss the merits of shadow rate models in capturing market expectations of the future path of monetary policy, but rather to illustrate that ${\mathbb L}$-forecasts can be close to risk-neutral ${\mathbb Q}$-forecasts and lead one far away from ${\mathbb P}$-forecasts, the point also made in a very different set of numerical examples in \citet{borovicka_2014mis}.

\begin{table}[H]
\begin{center}
\begin{tabular}{|c | c c| }
\hline
&  Median & Mean \\\hline
$\mathbb{P}$ & $0.33$ & $1.07$\\
$\mathbb{Q}$ & $0.17$ & $0.34$\\
$\mathbb{L}$ & $0.16$ & $0.32$ \\
\hline
\end{tabular}
\quad
\begin{tabular}{|c | c c| }
\hline
&  Median & Mean \\\hline
$\mathbb{P}$ & $2.13$ & $2.83$\\
$\mathbb{Q}$ & $1.34$ & $1.47$\\
$\mathbb{L}$ & $1.32$ & $1.46$ \\
\hline
\end{tabular}
\end{center}
\caption{Median and mean of the distribution of ZIRP lift-off time under $\mathbb{P}$, $\mathbb{Q}$ and $\mathbb{L}$ as of Aug. 19, 2015 (left) and Dec. 30, 2011 (right)}
\label{rate_lift1}
\end{table}

\section{Concluding Remarks}

This paper has demonstrated that the martingale component in the long-term factorization of the stochastic discount factor (SDF) due to  \citet{alvarez_2005using} and \citet{hansen_2009} is highly volatile, produces a downward-sloping term structure of bond Sharpe ratios as a function of bond's maturity, and implies that the long bond is far from growth optimality.  In contrast, the long forward probabilities forecast a generally upward sloping term structure of bond Sharpe ratios that starts from zero for short-term bonds and increases towards the Sharpe ratio of the long bond, and implies that the long bond is growth optimal.
Our empirical findings show that the assumption of transition independence of the SDF and degeneracy of the martingale component in its long-term factorization is implausible in the US Treasury bond market.

Our results in this paper are based on estimating a particular DTSM. We chose this DTSM as a representative model from the literature on term structure models respecting the zero bound.
While choosing a different model specification (in particular adding a third factors) would result in some quantitative differences, our
qualitative conclusions that the martingale component is highly volatile and produces the generally downward-sloping term structure of bond Sharpe ratios, as opposed to the long forward probability forecast of generally upward sloping term structure of bond Sharpe ratios, are robust to choosing a particular model specification.

\bibliography{mybib7}

\end{document}